\documentclass[format=acmsmall, nonacm, authorversion=true]{acmart}
\usepackage[utf8]{inputenc}
\usepackage{amsmath}
\usepackage{amsthm}
\usepackage[capitalise]{cleveref}

\makeatletter
\newtheorem*{rep@theorem}{\rep@title}
\newcommand{\newreptheorem}[2]{%
\newenvironment{rep#1}[1]{%
 \def\rep@title{#2 \ref{##1}}%
 \begin{rep@theorem}}%
 {\end{rep@theorem}}}
\makeatother

\newtheorem{theorem}{Theorem}
\newreptheorem{theorem}{Theorem}
\newtheorem{lemma}{Lemma}
\newtheorem{definition}{Definition}

\setcitestyle{acmnumeric}

\newcommand{\eat}[1]{}

\title{Incentive Compatible Queues Without Money}% \\ Paper \#91}
\author{Isaac Grosof}
\orcid{0000-0001-6205-8652}
\affiliation{%
    \institution{Carnegie Mellon University}
    \department{Computer Science Department}
    \city{Pittsburgh}
    \state{PA}
    \country{USA}
}
\email{igrosof@cs.cmu.edu}
\author{Michael Mitzenmacher}
\orcid{0000-0001-5430-5457}
\affiliation{%
    \institution{Harvard University}
    \department{School of Engineering and Applied Sciences}
    \city{Cambridge}
    \state{MA}
    \country{USA}
}
\begin{document}

\begin{abstract}
For job scheduling systems, where jobs require some amount of processing and then leave the system, it is natural for each user to provide an estimate of their job's time requirement in order to aid the scheduler.  However, if there is no incentive mechanism for truthfulness, each user will be motivated to provide estimates that give their job precedence in the schedule, so that the job completes as early as possible.

We examine how to make such scheduling systems incentive compatible, without using monetary charges, under a natural queueing theory framework.  In our setup, each user has an estimate of their job's running time, but it is possible for this estimate to be incorrect.  We examine scheduling policies where if a job exceeds its estimate, it is with some probability ``punished'' and re-scheduled after other jobs, to disincentivize underestimates of job times.  However, because user estimates may be incorrect (without any malicious intent), excessive punishment may incentivize users to overestimate their job times, which leads to less efficient scheduling.  We describe two natural scheduling policies, BlindTrust and MeasuredTrust.  We show that, for both of these policies, given the parameters of the system, we can efficiently determine the set of punishment probabilities that are incentive compatible, in that users are incentivized to provide their actual estimate of the job time.
Moreover, we prove for MeasuredTrust that in the limit as estimates converge to perfect accuracy,
the range of punishment probabilities that are incentive compatible converges to $[0,1]$.
Our formalism establishes a framework for studying further queue-based scheduling problems where job time estimates from users are utilized, and the system needs to incentivize truthful reporting of estimates.  

\end{abstract}

\maketitle

%\newpage

\section{Introduction}

Many job scheduling systems ask for estimates from users of how long their jobs will take \cite{bailey2004user,delgado2018kairos,ilyushkin2018impact,kettimuthu2005selective,tsafrir2007backfilling}.
% Add cites
% Analyzing and adjusting user runtime estimates to improve job scheduling on the Blue Gene
% slurm page:  https://slurm.schedmd.com/sbatch.html 
% time field
As a modern example, the open source SLURM workload manager for clusters (generally) requires users to provide a time limit for their job;
jobs that exceed their specified time are killed,
and jobs with smaller time requirements receive preference in their scheduling priority
\cite{slurm}.
Having one's job killed if it exceeds a provided time estimate naturally encourages estimates more conservative than one's true beliefs,
leading to less efficient scheduling.  

Surprisingly, there appears to be little previous work on theoretical analyses for 
incentive-compatible
scheduling systems with such voluntary estimates, even for the basic case of simple queues.  In particular, here we explore incentive-compatible queueing systems that do not use money or prices, which have received negligible study.  Instead, users may be ``punished'' by being moved further back in the queue when they exceed their estimates, incentivizing accuracy. (We discuss related work using money in \cref{sec:money}.)

We focus on the following natural problem formulation, based on a variation of the standard M/G/1 queue\footnote{An M/G/1 queue has Poisson (memoryless) arrivals (denoted by the M), general i.i.d. service times governed by a common distribution (denoted by the G), and a single server (denoted by the 1).}:  customers arrive to a single-server queue, according to a Poisson arrival process.  Using Poisson arrivals allows us to use 
standard queueing theoretic frameworks.
We focus on the expected time a job spends in the queue in equilibrium, in a setting where a user is unaware of the actual state of the queue when their job arrives.

Each job has one of a finite number of true types $\{1,2,\ldots,n\}$, which determines their service time requirements;  a customer of true type $i$ has service time $z_i$, where we let $z_1 < z_2 < \ldots < z_n$.  Each job also has an internal estimate of its time;  that is, each job may have a belief that it is of type $j$.  The internal estimate $j$ of a job need not be equal to its true type $i$.  Finally, each job has a declared estimate $k$, which the user provides to the system, and which may differ from both its true type and its internal estimate.
We focus on a discrete setting for simplicity.

If all declared estimates were equal to the true type, the natural scheduling strategy, which minimizes the expected time a job spends in the system, would be Shortest Job First%
\footnote{
    More specifically, Shortest Remaining Processing Time first,
    which adjusts for partially complete jobs.
}
(where ties, for jobs of the same type, could be broken by first-come first-served).  Under such a scheduling strategy, if declared estimates are used and there is no punishment for an incorrect estimate, users are incentivized to report that their jobs are type 1, to minimize their own time waiting for service.  That is, users are incentivized to lie.  On the other hand, as mentioned above, severe enough scheduling strategies, such killing jobs that exceed their declared estimates,
or automatically moving such jobs to the end of the queue, may encourage users to declare estimates larger than their internal estimate, to avoid a severe penalty in case their internal estimate is incorrect.

We seek to understand scheduling strategies which are incentive compatible, in the sense that every user's best option is to declare their internal estimate, and the conditions under which such strategies exist.  We emphasize here that we consider the {\em stochastic} version of this problem, where users do not know the state of the queue, and thus base their reporting decision on the expected time they would spend in the system in equilibrium, rather than a worst-case version of the problem where users have more information (such as the state of the system, or even the arrival times of future jobs).     

We consider two scheduling variants, which are described more formally in \cref{sec:model}. BlindTrust schedules jobs according to their estimated time, but if a job exceeds its estimate, with a fixed probability $b$ it is given the lowest possible priority, below all other jobs that have not exceeded their estimate;  with all remaining probability, the job continues according to its time estimate.  (Jobs with the lowest priority are served in order of their initial arrival time, i.e. First Come First Served.)  MeasuredTrust similarly gives a job that exceeds its initial estimate the lowest priority with probability $b$, but instead of leaving the job at that priority with the remaining probability, MeasuredTrust lowers the priority of the job to the next priority level, and continues to lower its priority when the job exceeds the time associated with its current priority type.

In \cref{sec:results},
we state our main results.
In \cref{thm:alg-bounce}, we prove that the range of $b$ values for both policies that incentivize truthful reporting of estimates by users can be calculated efficiently in our model.
In \cref{thm:single-interval}, we prove that for any setting,
the set of parameters $b$ for which MeasuredTrust
is incentive compatible
forms a single connected interval.
Finally, in \cref{thm:small-noise},
we prove that as user estimates converge to perfect accuracy,
the range of $b$ values for which
MeasuredTrust becomes incentive compatible converges to the full interval $[0,1]$.

Finally, in \cref{sec:empirical}, we empirically evaluate MeasuredTrust and BlindTrust,
showing that scheduling with estimates can lead to significant gains over blind policies such as First Come First Served.

\section{Related Work}

Our work utilizes a queueing theory framework.  
For an introduction to queueing theory generally, we recommend the recent text \cite{harchol2013performance}.  

\subsection{Incentive compatible queues with money}
\label{sec:money}

Prior work on incentive compatible queues has focused on paid priority mechanisms;  for example, customers choose and pay for a class of service, which determines their priority in the queue, based on the value of their waiting time (see e.g. \cite{afeche2013incentive,gupta2000extracting,lederer1997pricing,mendelson1990optimal}).  From the point of view of the queue operator, the goal might be to develop a priority-based scheme that maximizes revenue.  

Our work takes a different tack, avoiding pricing, as might be appropriate for a cooperatively shared resource, such as a supercomputer in a university, or an internally used data center within a company. Instead, we use the simple mechanism of preempting  a job that exceeds its declared estimate while initially ordering jobs according to their declared estimates to encourage truthful reporting.   Also, we are not aware of work on incentive compatible queues that considers that customers may mis-estimate their job properties, such as the required service time, as we do here.

\subsection{Queueing with estimates}

Queues using estimated service times rather than exact service times have been studied both empirically and analytically \cite{DCM, dellamico2019scheduling, mitz2019scheduling, mitzenmacher2019supermarket, mitzenmacher2021queues, scully2018, scully2021uniform, WiermanNuyens}.  These works have focused on queue performance when using various scheduling policies with estimates, such as Shortest Predicted Remaining Processing Time or Shortest Predicted Job First (in place of the policies Shortest Remaining Processing Time and Shortest Job First, which are well-studied policies with exact information).  These works generally assume the estimates are provided by an external process, such as a machine learning algorithm, rather than from user agents themselves, and therefore do not consider the issue of incentivizing accurate estimates.

\subsection{Queueing background: SOAP}

Our work makes use of the fact that the expected time a job spends in the system (in equilibrium) can be exactly calculated for systems that use ranks to determine system priority, where the rank can depend on the type of the job as well as the time the job has spent in service.  Recent work has derived a uniform framework for analyzing age-based policies, referred to as SOAP (Schedule Ordered by Age-based Priority) \cite{scully2018soap}, which we use in \cref{sec:formulas} to derive mean response time formulas for our policies.

Our policies are somewhat similar to a previously considered class of policies known as
\emph{multilevel processor sharing with FCFS tiebreaking} \cite{kleinrock1976},
but we rely on the additional generality of SOAP to cleanly handle
job size estimates and our punishment parameter $b$.

\section{Definitions and Notation}
\label{sec:model}
We provide the definitions and notation we use throughout, and formally define our model.  

Jobs have two inherent properties: a {\em size} $s$ and a {\em size estimate} $r$.
Both the size and the estimate are elements of a finite set $Z$ with $n$ elements,
\begin{align*}
    &Z := {z_1, z_2, \ldots, z_n} \\
    &z_1 < z_2 < \ldots < z_n \\
    & s = z_i \in Z, r = z_j \in Z.
\end{align*}
The estimate is accurate if $i = j$, and otherwise it is inaccurate.
We focus on a finite set of sizes and estimates for simplicity.
In particular, a continuous model would make stating and proving results analogous
to our results in \cref{sec:results} considerably more complicated.

Each job is associated with a user;  we assume each user is associated with a single job (avoiding issues of a user controlling multiple jobs).    
After a job is generated, the user sees the size estimate $z_j$
and chooses a {\em declared estimate} $z_k \in Z$ to provide to the scheduler.
For clarity, we call the original size estimate $z_j$ the {\em internal estimate}
to differentiate it from $z_k$. We say that
a user is {\em honest} if their declared estimate $z_k$ matches their internal estimate $z_j$,
and \textit{lying} otherwise.

Jobs arrive according to a Poisson process with rate $\lambda$,
and are given a declared estimate by a user immediately and irrevocably.
We consider each job to be assigned a declared estimate by a different user,
and users to make decisions without communicating or collaborating.

Job sizes $z_i$ and internal estimates $z_j$ are sampled i.i.d. from some joint distribution $(S, R)$,
both supported on $Z$. We assume that the users and scheduler have observed many arrivals,
so the joint distribution is known to all parties.
The joint distribution can be represented as a matrix $M$,
where $M_{ij} = P(S = z_i, R = z_j)$.
We also write $S_i = P(S = z_i)$ and $R_j = P(R = z_j)$ for the marginals of this distribution.  The {\em load} $\rho = \lambda E[S]$ denotes the fraction of time the server is occupied.
We assume that $\rho < 1$ to ensure stability.

The scheduler decides whether to serve a job based on its declared estimate $z_k$
and its {\em age} $a$, the amount of service it has received so far.
Note that the scheduler does not have access to either the internal estimate $z_j$,
or the true size $z_i$.
A job completes when its age reaches its size $z_i$.
We assume that the scheduler can preempt jobs at any time without loss of work.

We now define several random variables that correspond to different characterizations of {\em response time} that we use in our analysis.
A job's response time is the time from when the job arrives to when the job completes.

First, suppose that an identified job has true size $z_i$ and declared estimate $z_k$.
Suppose also that all other jobs have honest declared estimates.
Suppose that the identified job arrives at a generic moment in time, when the system state is in its stationary state.
Let $U_{ik}$ denote the random variable corresponding to the response time of the identified job.
Note that we implicitly make use of the PASTA principle
\cite{harchol2013performance},
which states that Poisson arrivals
see the stationary state of the system.

Next, suppose that an identified job has internal estimate $z_j$,
keeping all other assumptions the same.
Let $T_{jk}$ denote the random variable corresponding to the response time of this identified job.
Note that
\vspace{-.1in}
\begin{align*}
    T_{jk} = \frac{1}{R_j} \sum_{i \in [n]} M_{ij} U_{ik}.
\end{align*}

Finally, to denote the overall response time of a scheduling policy,
under the assumption that all users are honest,
we use the random variable $T$.
Note that
\begin{align*}
    T = \sum_{j \in [n]} R_j T_{jj}.
\end{align*}

We are interested in designing scheduling policies with two properties:
\emph{incentive compatibility} and \emph{social benefit}.

\begin{definition}
    A scheduling policy is \emph{incentive compatible}
    if each user is incentivized to be honest, under the assumption that
    all other users are honest.
    In our setting, for all internal estimates $z_j$ and all declared estimates $z_k$, this implies
    \begin{align*}
        E[T_{jj}] \le E[T_{jk}].
    \end{align*}
\end{definition}

\begin{definition}
    A scheduling policy is \emph{socially beneficial} with respect to some baseline estimate-blind policy $\pi$
    if the scheduling policy produces a better societal outcome when all users are honest
    than the baseline policy $\pi$.
    In our setting, using $T^{\pi}$ for the equilibrium response time under $\pi$, this implies
    \begin{align*}
        E[T] \le E[T^{\pi}].
    \end{align*}
    Common baseline policies might include First-Come-First-Served (FCFS), Foreground-Background (FB) \cite{nuyens2008foreground},
    or the Gittins policy \cite{gittins2011multi}.
\end{definition}

Throughout the paper, we will typically use the indices $i$, $j$, $k$ in the following fashion:
\begin{itemize}
    \item $i$ is the index of a job's true size. We use this with $U$,
    the response time distribution of a job of a given true size and declared estimate.
    \item $j$ is the index of a job's internal estimate, specifying a distribution over true sizes given by a column of the
    size-estimate matrix $M$.
    We use this with $T$, the response time distribution of a job of a given internal estimate and declared estimate.
    \item $k$ is the index of a job's declared estimate,
    the estimate of size stated by the user to the system.
    We use $k$ with both $T$ and $U$.
\end{itemize}

We now define our two scheduling policies of interest:
\emph{MeasuredTrust} and \emph{BlindTrust}.
Each policy is parameterized by a \emph{punishment parameter} $b$.

MeasuredTrust has $n+1$ priority classes.
MeasuredTrust preemptively serves the job of lowest numbered priority class in the system,
and serves jobs in arrival order within a given priority class.
MeasuredTrust initially places a job with declared estimate $z_j$ into priority class $j$.
However, if a job in priority class $j$ reaches age $z_j$ without completing,
MeasuredTrust deprioritizes it in one of two ways.
To choose how to deprioritize the job, the scheduler flips a coin with probability $b$ of coming up heads.
If the coin comes up heads,
MeasuredTrust moves the job to priority class $n+1$,
the highest numbered priority class (with lowest actual priority).
Otherwise,
MeasuredTrust moves the job to priority class $j+1$.
If the job does not complete at age $z_{j+1}$,
MeasuredTrust moves it to class $j+2$ without reflipping the coin,
and similarly continues moving it to higher numbered priority classes as it ages beyond larger values $z \in Z$ until it completes.

BlindTrust is defined similarly to MeasuredTrust, with a single change.
If the probability $b$ coin flip comes up tails,
BlindTrust leaves the job in class $j$ until completion.  That is, the job either stays in its initial priority class based on its declared estimate, or goes directly to the lowest priority class.

In our empirical results in \cref{sec:empirical},
we find that MeasuredTrust achieves incentive compatibility under noisier conditions
and across a wider range of punishment parameters $b$ in our example setting,
while BlindTrust achieves slightly lower mean response time
when it can achieve incentive compatibility in our example setting.  As such we cannot say that one strategy is strictly better than the other;  it may depend both on the system parameters, as well as on the  desired system goals.  

\subsection{Example}

Before getting to our results, we provide a brief example to help illustrate our setting.  Let the size and estimate set $Z = [1, 2, 3]$, and let the size to internal estimate correlation matrix $M$
be
\begin{align*}
    \begin{bmatrix}
        0.425 & 0.03 & 0.01\\
        0.05 & 0.255 & 0.02\\
        0.025 & 0.015 & 0.17
    \end{bmatrix}
\end{align*}

For instance, the entry $M_{12} = 0.03$ states
that there is a $0.03$ probability of a job having
size $z_1 = 1$ and internal estimate $z_2 = 2$.

We can likewise look at marginal distributions.
Summing the third column, we find that $R_3$,
the probability of a job having internal estimate $3$,
is $0.01 + 0.02 + 0.17 = 0.2$.
Of jobs with internal estimate 3, $0.17/0.2 = 85\%$ also have true size 3.

Consider the case where the arrival rate is $\lambda = 0.5$ and the scheduling policy is MeasuredTrust.
Using the formulas we provide in \cref{sec:formulas},
we can calculate $E[T_{jk}]$ for any given punishment probability $b$.  Using this, we can find the range of $b$ values for which MeasuredTrust is incentive compatible.  (We consider $b$ to two decimal places for convenience.)

For $b \in [0, 0.14]$,
MeasuredTrust is not incentive compatible,
because certain users gain from not being honest.
In particular, $E[T_{33}] > E[T_{31}]$
for $b$ in this range.  That is, a user with an internal estimate of $3$ would obtain a lower expected time in service by declaring a lower estimate of $1$ due to the potential gain if their internal estimate is incorrect and the low probability of punishment.

For $b \in [0.71, 1]$,
MeasuredTrust is again not incentive compatible,
because certain users gain from not being honest, but in this case these users are motivated to provide a higher estimate. In particular, $E[T_{11}] > E[T_{12}]$
for $b$ in this range.
Users with internal estimate $1$ are incentivized to declare an estimate of $2$ to avoid being punished even if they are being honest but the internal estimate is inaccurate.

For $b \in [0.15, 0.70]$,
MeasuredTrust is incentive compatible,
meaning that all users are incentivized to be honest.  This is found by checking three inequalities:
\eat{
\begin{align*}
E[T_{11}] \leq \min(E[T_{12}],E[T_{13}]) \\
E[T_{22}] \leq \min(E[T_{21}],E[T_{23}]) \\
E[T_{33}] \leq \min(E[T_{31}],E[T_{32}])
\end{align*}
}
\begin{align*}
E[T_{11}] \leq \min(E[T_{12}],E[T_{13}]) ; \,\,
E[T_{22}] \leq \min(E[T_{21}],E[T_{23}]) ; \,\,
E[T_{33}] \leq \min(E[T_{31}],E[T_{32}]).
\end{align*}

For $b \in [0.15, 0.70]$,
MeasuredTrust is also always socially beneficial relative
to FCFS.
The overall mean response time for MeasuredTrust goes from $7.199$ at $b=0.15$ down to $7.000$ at $b=0.43$ and back up $7.276$ at $b=0.71$,
while $E[T^{FCFS}] = 8.912$ with these parameters.

For BlindTrust, we have a narrower range of incentive compatibility.
For $b \in [0, 0.80]$, $E[T_{33}] > E[T_{31}]$,
while for $b \in [0.86, 1]$, $E[T_{11}] > E[T_{12}]$.
For $b \in [0.81, 0.85]$, BlindTrust is incentive compatible.
For $b \in [0.81, 0.85]$, BlindTrust is also socially beneficial relative to FCFS.
The overall expected response time for BlindTrust
goes from $6.553$ at $b=0.81$ up to $6.792$ at $b=0.85$,
all smaller than $E[T^{FCFS}] = 8.912$.

Note that in this example, despite its narrower range of incentive compatibility,
BlindTrust achieves superior overall mean response time
to MeasuredTrust when setting $b$ optimally.

\section{Results}
\label{sec:results}

We here first describe our results at a higher level.  The proofs require various technical derivations from queueing theoretical analysis, which we provide in Section~\ref{sec:formulas}.  Armed with the needed formulas, we provide full proofs in Section~\ref{sec:proofs}.

Our first result says that for both MeasuredTrust and BlindTrust, there are polynomial time algorithms to determine the set of punishment probabilities that are incentive compatible, and for which the resulting expected is socially beneficial.  (Recall that socially beneficial means better in expectation than a baseline policy.)
\begin{theorem}[Poly-time Incentive Checking Algorithm]
    \label{thm:alg-bounce}
    For a given size-estimate matrix $M$ and a given arrival rate $\lambda$,
    there is poly time algorithm to find all punishment probabilities $b$ such that MeasuredTrust (BlindTrust)
    is incentive compatible and socially beneficial.
\end{theorem}

Our next two theorems focus on MeasuredTrust, which is more often incentive compatible.
Our next result refines the previous result,
showing that in fact the set of punishment probabilities
that are incentive compatible is actually an interval.
Knowing this can lead to more efficient algorithms to find this interval.
\begin{theorem}[Single Interval]
    \label{thm:single-interval}
    For a given size-estimate matrix $M$ and a given arrival rate $\lambda$,
    there exist cutoffs $b^*_{low}, b^*_{high}$
    such that MeasuredTrust is incentive compatible
    if and only if the punishment probability $b$ is in the interval $[b^*_{low}, b^*_{high}]$.
    (Note that for some scenarios, this interval may be empty.)
\end{theorem}

Our final result characterizes MeasuredTrust further.  We show that as the user estimates converge toward perfect accuracy, the interval of incentive compatible punishment probabilities converges to $[0,1]$.
\begin{theorem}[Small Noise]
    \label{thm:small-noise}
    For a given arrival rate $\lambda$ and true size distribution $S$,
    there is some threshold $\epsilon$
    such that for any size-estimate matrix $M$
    compatible with $S$ such that $|M - D_S|_\infty < \epsilon$,
    where $D_S$ is the diagonal, perfectly accurate size-estimate matrix,
    there must exist a punishment probability $b$
    such that Measured Trust is incentive compatible.
    Moreover, as $\epsilon \to 0$,
    the interval of $b$ for which MeasuredTrust is incentive compatible
    converges to $[0, 1]$.
\end{theorem}

\section{Formulas}
\label{sec:formulas}

To analyze the MeasuredTrust and BlindTrust policies,
we make use of the recently developed SOAP framework \cite{scully2018soap}
to derive exact formulas for mean response time.

\subsection{SOAP Background}
The SOAP framework can be used to analyze
scheduling policies for M/G/1 queues (Poisson arrivals, general service distribution, single server) that can be expressed in terms of \emph{rank functions}.

A rank-function-based scheduling policy determines which job to serve by computing the \emph{rank} of each job, and always serving the job with least rank (preempting jobs as needed).
If there is a tie for lowest rank, tiebreaking is First Come First Served (FCFS), based on order of arrival to the system.
The rank function determines the rank of each job,
based on its age, class, and other static qualities.
In our case, 
% we are computing $U_{ik}$,
a job's characteristics are
its age $a$,
its declared estimate index $k$,
and an indicator variable $I_b$, which specifies whether the job will be punished by moving it to the back of the queue (so that it will have the worst priority).  
Specifically, we imagine that the probabilistic decision of whether to punish the job by moving it to the back if it exceeds its declared estimate is made when the job arrives, and we let $I_b$ be the result of this decision.

Given a rank function,
we can use the SOAP approach to write down a formula for mean response time \cite{scully2018soap}.
To do so, we first define the ``relevant size'' $S_{\le r}$ for each rank $r$,
a random variable denoting the amount of time for which a general (honest) job
receives service while the job's rank is less than or equal to $r$.
We define $S_{< r}$ similarly.
In particular, we define $S_{\le r}$
for a general honest job,
because we are operating on the assumption that all jobs provide honest estimates,
except for potentially the job being focused on.
With $S_{\le r}$ defined, we also write $\rho_{\le r} = \lambda E[S_{\le r}]$
to denote the load of rank $\le r$, and $\rho_{< r}$ similarly.

Next, we define the \emph{worst future rank} for each job.
In our setting, this is simple to do,
because as we shall see in \cref{sec:measured-trust-formula} and
\cref{sec:blind-trust-formula},
our rank functions are weakly monotonically increasing
as each job receives service.
As a result, a job's worst future rank is simply its final rank.
We write this worst future rank as $w(i, k, I_b)$.

Now, using SOAP \cite{scully2018soap}, we can write down the formula for a job's mean response time, where we let $w = w(i,k,I_b)$:
\begin{align}
    \label{eq:generic-soap}
    E[U_{i,k,I_b}] &= \frac{\lambda E[S^2_{\le w}]}{2(1-\rho_{< w})(1-\rho_{\le w})}
    + \frac{z_i}{1-\rho_{< w}}
\end{align}
Note that because our rank functions are weakly monotonically increasing,
several terms from the fully general SOAP formula \cite[Theorem~5.5]{scully2018soap} simplify or vanish.

Here we see that the expected time a job spends in the system depends on its worst future rank,
as well as the relevant size variables $S_{\le r}$. 
We now examine the behavior of the worst future rank and of the relevant size variables for our two scheduling algorithms, and in particular determine how these expressions depend on the punishment parameter $b$.

\subsection{MeasuredTrust}
\label{sec:measured-trust-formula}

\begin{lemma}
\label{lem:measured-trust-formula}
    For the MeasuredTrust policy,
    mean response time for a job of true size $i$
    and declared estimate $k$ is
\begin{align}
    \label{eq:measured-trust-safe}
    \forall i \le k, \quad E[U_{ik}] &=
    \frac{\lambda E[S^2_{\le k}]}{2(1-\rho_{<k})(1-\rho_{\le k})}
    + \frac{z_i}{1-\rho_{<k}} \\
    \label{eq:measured-trust-unsafe}
    \forall i > k, \quad E[U_{ik}] &= bE[U_{ik,I_b=1}] + (1-b)E[U_{ik,I_b=0}] \\
    \label{eq:measured-trust-bounced}
    \forall i > k, \quad E[U_{ik,I_b=1}] &=
    \frac{\lambda E[S^2]}{2(1-\rho_{<n+1})(1-\rho)}
    + \frac{z_i}{1-\rho} \\
    \label{eq:measured-trust-unbounced}
    \forall i > k, \quad E[U_{ik,I_b=0}] &=
    \frac{\lambda E[S^2_{\le i}]}{2(1-\rho_{<i})(1-\rho_{\le i})}
    + \frac{z_i}{1-\rho_{< i}}
\end{align}
where $S_{\le i}$ and $\rho_{\le i}$ are defined below in terms of $\lambda, S,$ and $M$.
\end{lemma}
\begin{proof}

To analyze MeasuredTrust using SOAP,
we first write MeasuredTrust's rank function,
mapping a job's age $a$,
its declared estimate index $k$,
and its punishment indicator $I_b$
to its rank, where the lowest rank is served,
with FCFS tiebreaking.  Our ranks are the integers $[1, 2, \ldots, n, n+1]$.

Recall from \cref{sec:model}
how MeasuredTrust works,
using the terminology of ranks.
A job is initially served with rank equal to its declared estimate $k$.
If age $z_k$ is exceeded,
the declared estimate was wrong.
Now, a coin is flipped with probability $b$.
If heads, the job is punished by increasing its rank
to $n+1$, where its rank stays until it completes.
If tails, the job's rank is only increased to $k+1$.
If the job does not finish at age $z_{k+1}$,
its rank is increased to $k+2$, and so on until it completes.

The MeasuredTrust rank function $r_{MT}$ is:
\begin{align*}
    r_{MT}(a, k, I_b) = \begin{cases}
        a < z_k & k \\
        a \ge z_k \land I_b = 1 & n+1 \\
        a \ge z_k \land I_b = 0 & \arg\min_{\ell \in [n]} \{a < z_{\ell}\}
    \end{cases}
\end{align*}

The first case covers jobs with age below their estimates,
the second case covers punished jobs,
and the third case covers unpunished jobs with age above their estimates. As previously stated, $r_{MT}$ is weakly monotonically increasing with respect to age $a$.

Now, we can specify the random variable $S_{\le \ell}$
for each rank $\ell \in [1, \ldots, n+1]$.
To do so, let us first define $S_{ij, \le \ell}$ to be the 
% conditional
service at ranks $\le \ell$ for honest jobs with true size index $i$
and internal estimate index $j$.
Then $S_{\le \ell}$ is simply
%\vspace{-0.05in}
\begin{align*}
    S_{\le \ell} = \sum_{i \in [n], j \in [n]} M_{ij} S_{ij, \le \ell}.
\end{align*}

Now, it remains to specify $S_{ij, \le \ell}$.
For $\ell \le n$:
\vspace{-0.1in}
\begin{align}
    \label{eq:relevant-size}
    S_{ij, \le \ell} = \begin{cases}
        j > \ell & 0 \\
        i \le j \le \ell & z_i \\
        j < i \land j \le \ell \land I_b = 1 & z_j \\
        j < i \le \ell \land I_b = 0 & z_i  \\
        j \le \ell < i \land I_b = 0 & z_\ell
    \end{cases}
\end{align}
Note that $I_b$ here is a Bernoulli distributed random variable
with parameter $b$.

To understand $S_{ij,\le \ell}$, let us examine some cases.
The third case, $j < i \land j \le \ell \land I_b = 1$,
covers the scenario where a job exceeds its estimate after $z_j$ service,
and is punished, immediately increasing its rank above $\ell$.
On the other hand, the fifth case, $j \le \ell < i \land I_b = 0$
covers the case where a job exceeds its estimate,
is not punished,
and only reaches a rank above $\ell$ after $z_\ell$ service.

As a special case, $S_{n+1} = S$.
We can similarly define $S_{< \ell} = S_{\le \ell - 1}$.
As a special case, $S_{< 1} = 0$.

Next, we need to write down the final-rank function $w(i, k, I_b)$:
\begin{align*}
    w(i, k, I_b) = \begin{cases}
        i \le k & k \\
        i > k \land I_b = 1 & n+1 \\
        i > k \land I_b = 0 & i
    \end{cases}
\end{align*}

Now \cref{lem:measured-trust-formula}
follows from specializing the generic formula \eqref{eq:generic-soap}:

\begin{align*}
    \forall i \le k, \quad E[U_{ik}] &=
    \frac{\lambda E[S^2_{\le k}]}{2(1-\rho_{<k})(1-\rho_{\le k})}
    + \frac{z_i}{1-\rho_{<k}} \\
    \forall i > k, \quad E[U_{ik}] &= bE[U_{ik,I_b=1}] + (1-b)E[U_{ik,I_b=0}] \\
    \forall i > k, \quad E[U_{ik,I_b=1}] &=
    \frac{\lambda E[S^2]}{2(1-\rho_{<n+1})(1-\rho)}
    + \frac{z_i}{1-\rho} \\
    \forall i > k, \quad E[U_{ik,I_b=0}] &=
    \frac{\lambda E[S^2_{\le i}]}{2(1-\rho_{<i})(1-\rho_{\le i})}
    + \frac{z_i}{1-\rho_{< i}}
    \qedhere
\end{align*}
\end{proof}

It is helpful to understand how our final expressions depend on $b$.
Looking at \eqref{eq:relevant-size},
the only dependency of $S_{ij,\le \ell}$ on $b$ is through $I_b$,
meaning that $S_{ij,\le \ell}$ depends linearly on $b$.
Likewise, $S_{\le \ell}$ depends linearly on $b$,
and so do $E[S^2_{\le \ell}]$ and $\rho_{\le \ell}$.
As a result, in each case $E[U_{ik}]$ is a constant-degree rational polynomial in $b$.

\subsection{BlindTrust}
\label{sec:blind-trust-formula}

\begin{lemma}
\label{lem:blind-trust-formula}
For the BlindTrust policy,
mean response time for a job of true size $i$ and declared size estimate $k$ is
\begin{align*}
    \forall i \le k, \quad E[U^{BT}_{ik}] &=
    \frac{\lambda E[(S^{BT}_{\le k})^2]}{2(1-\rho^{BT}_{<k})(1-\rho^{BT}_{\le k})}
    + \frac{z_i}{1-\rho^{BT}_{<k}} \\
    \forall i > k, \quad E[U^{BT}_{ik}] &= bE[U^{BT}_{ik,I_b=1}] + (1-b)E[U^{BT}_{ik,I_b=0}] \\
    \forall i > k, \quad E[U^{BT}_{ik,I_b=1}] &=
    \frac{\lambda E[S^2]}{2(1-\rho^{BT}_{<n+1})(1-\rho)}
    + \frac{z_i}{1-\rho} \\
    \forall i > k, \quad E[U^{BT}_{ik,I_b=0}] &=
    \frac{\lambda E[(S^{BT}_{\le k})^2]}{2(1-\rho^{BT}_{<k})(1-\rho^{BT}_{\le k})}
    + \frac{z_i}{1-\rho^{BT}_{<k}} \\
\end{align*}
where $S^{BT}_{\le i}$ and $\rho^{BT}_{\le i}$ are defined below in terms of $\lambda, S,$ and $M$.
\end{lemma}
\begin{proof}

To analyze BlindTrust using SOAP,
we follow the same steps as for MeasuredTrust in \cref{sec:measured-trust-formula}.
Again, our ranks are the integers $[1, 2, \ldots, n, n+1]$.

The BlindTrust rank function $r_{BT}$ is
\begin{align*}
    r_{BT}(a, k, I_b) = \begin{cases}
        a < z_k & k \\
        a \ge z_k \land I_b = 1 & n+1 \\
        a \ge z_k \land I_b = 0 & k
    \end{cases}
\end{align*}
The only difference from $r_{MT}$ comes in the $I_b = 0$ case.
Note that $r_{BT}$ is also weakly monotonically increasing with age $a$.

Next, we specify $S^{BT}_{\le \ell}$ and $S^{BT}_{ij,\le \ell}$.
When the superscript is omitted, we refer to MeasuredTrust by default.
\begin{align*}
    S^{BT}_{\le \ell} &= \sum_{i\in[n],j\in[n]} M_{ij}S^{BT}_{ij,\le \ell} \\
    \forall \ell \le n \quad S^{BT}_{ij,\le \ell} &= \begin{cases}
        j > \ell & 0 \\
        i \le j \le \ell & z_i \\
        j < i \land j \le \ell \land I_b = 1 & z_j \\
        j < i \land j \le \ell \land I_b = 0 & z_i
    \end{cases}
\end{align*}
The difference from $r_{MT}$ comes in the $j < \ell < i \land I_b = 0$ case.
Again, as a special case, $S^{BT}_{n+1} = S, S^{BT}_{<\ell} = S^{BT}_{\le \ell-1},
S_{<1}=0$.

Next, we specify the final-rank function $w^{BT}(i, k, I_b)$:
\begin{align*}
w^{BT}(i, k, I_b) =
\begin{cases}
    i \le k & k \\
    i > k \land I_b = 1 & n+1 \\
    i > k \land I_b = 0 & k
\end{cases}
\end{align*}

Now \cref{lem:blind-trust-formula} follows from
specializing the generic formula \eqref{eq:generic-soap}:

\begin{align*}
    \forall i \le k, \quad E[U^{BT}_{ik}] &=
    \frac{\lambda E[(S^{BT}_{\le k})^2]}{2(1-\rho^{BT}_{<k})(1-\rho^{BT}_{\le k})}
    + \frac{z_i}{1-\rho^{BT}_{<k}} \\
    \forall i > k, \quad E[U^{BT}_{ik}] &= bE[U^{BT}_{ik,I_b=1}] + (1-b)E[U^{BT}_{ik,I_b=0}] \\
    \forall i > k, \quad E[U^{BT}_{ik,I_b=1}] &=
    \frac{\lambda E[S^2]}{2(1-\rho^{BT}_{<n+1})(1-\rho)}
    + \frac{z_i}{1-\rho} \\
    \forall i > k, \quad E[U^{BT}_{ik,I_b=0}] &=
    \frac{\lambda E[(S^{BT}_{\le k})^2]}{2(1-\rho^{BT}_{<k})(1-\rho^{BT}_{\le k})}
    + \frac{z_i}{1-\rho^{BT}_{<k}}
    \qedhere
\end{align*}
\end{proof}

Note that the formulas for $E[U_{ik}]$ for $i \le k$ and for $E[U_{ik,I_b=0}]$ for $i > k$ are identical.  Also, here too, in each case $E[U^{BT}_{ik}]$ is a constant-degree rational polynomial in $b$.

\section{Proofs of results}
\label{sec:proofs}

\subsection{Finding Roots}

In what follows we assume we have ``black boxes'' for the following subtasks:
\begin{itemize}
    \item if there is at most one real root of a univariate polynomial of degree $n$ in the interval $[0,1]$, we can find it;
    \item we can find all the roots of a univariate polynomial of degree $n$ in the interval $[0,1]$.
\end{itemize}
Here when we say find a root of a univariate polynomial of degree $n$, we mean we can find the root to any desired precision in time polynomial in $n$ (that may depend on the precision).  The second task of course encompasses the first, but finding a single root is generally faster in theory and practice.  For instance, finding a single root can often be done efficiently using Newton's algorithm, while finding all the roots requires more complicated methods.  We do not concern ourselves with the exact complexity of root-finding here, which can depend on desired computational precision and the computational model, and is not our focus; we refer the reader to the various works on the subject (e.g.,
\cite{aberth1973iteration}, \cite{neff1996efficient}, \cite[Chapter 9]{vishnoi2021algorithms}).  

\subsection{Proof of Theorem 1}
\label{sec:poly-time}

We now turn to the proofs of the theorems.  The first is regarding finding the area of incentive compatibility in polynomial time.
% \subsection{Proof of Theorem 1}
% \label{sec:alg-bounce}

\begin{reptheorem}{thm:alg-bounce}[Poly-time Incentive Checking Algorithm]
    For a given size-estimate matrix $M$ and a given arrival rate $\lambda$,
    there is polynomial time algorithm to find all punishment probabilities $b$ such that MeasuredTrust (BlindTrust)
    is incentive compatible and socially beneficial.
\end{reptheorem}
\begin{proof}
% All of the quantities such as $E[S_{\le i}]$
%     are constant-degree polynomials in $b$, holding $M$ and $\lambda$ constant.
As noted in Section~\ref{sec:formulas}, the quantities $E[U_{ik}]$ are each constant-degree rational polynomials in $b$.  As each $E[T_{jk}]$ is a weighted mixture of $O(n)$ terms of the form $E[U_{ik}]$, each $E[T_{jk}]$ is a rational polynomial of degree $O(n)$ in $b$, where $n$ is the number of classes. Hence the expression $E[T_{jk}] -E[T_{jk'}]$ for $k \neq k'$ is a rational polynomial in $b$ of degree $O(n)$, and as such, after clearing the denominators from the equation 
$E[T_{jk}] -E[T_{jk'}] = 0$, 
we can find the values of $b$ that are roots of this equation in polynomial time.  Correspondingly, we can find the intersection of the intervals of $b$ where $E[T_{jj}] -E[T_{jk}] \geq 0$, over all the $O(n^2)$ values of $j$ and $k \neq j$, in polynomial time, which is the collection of punishment probabilities that are incentive compatible.  

Similarly, we can calculate the mean response time $M^*$ for a proposed scheduling scheme to compare against.  The condition
$\sum_j R_j E[T_{jj}] < M^*$ corresponds to the policy being socially beneficial.  The left-hand side is again a rational polynomial of degree $O(n)$, and hence in polynomial time we can also find the intervals of $b$ for which this equation holds.   
\end{proof}

\subsection{Proof of Theorem 2}
\label{sec:single-interval}

Theorem~\ref{thm:single-interval} shows that for MeasuredTrust the punishment probabilities that lead to incentive compatibility form an interval.  The proof allows for a more efficient polynomial time algorithm for finding the interval MeasuredTrust is incentive compatible over Theorem~\ref{thm:alg-bounce}, as it implies the equations $E[T_{jj}] -E[T_{jk}] \geq 0$ hold for intervals of $b$ the form $[0,b_{jk}]$ or $[b_{jk},1]$, and finding these $b_{jk}$ corresponds to finding a single root in the interval $[0,1]$ (instead of all of the roots of the corresponding polynomial equation).  

\begin{reptheorem}{thm:single-interval}[Single Interval]
    For a given size-estimate matrix $M$ and a given arrival rate $\lambda$,
    there exist cutoffs $b^*_{low}, b^*_{high}$
    such that MeasuredTrust is incentive compatible
    if and only if the bound probability $b$ is in the interval $[b^*_{low}, b^*_{high}]$.
    (Note that for some scenarios, this interval may be empty.)
\end{reptheorem}
\begin{proof}
    Compare $E[U_{ik}]$ to $E[U_{ik'}]$, where $k < k'$, and $i$ is a true size.
    Specifically compare their derivates with respect to $b$, $d/db$:
    \begin{align*}
        \frac{d}{db} E[U_{ik}] \stackrel{?}{<} \frac{d}{db} E[U_{ik'}]
    \end{align*}
    We will show that all the relative derivatives are ordered such that the smaller claimed estimate $k$
    is either less benefited or more harmed by increasing $b$:
    \begin{align*}
        \frac{d}{db} E[U_{ik}] \ge \frac{d}{db} E[U_{ik'}]
    \end{align*}
    
    As a result, the same holds for all $E[T_{jk}]$:
    \begin{align*}
        \frac{d}{db} E[T_{jk}] \ge \frac{d}{db} E[T_{jk'}]
    \end{align*}
    In particular, the difference $E[T_{jk}] - E[T_{jk'}]$
    is increasing everywhere (and continuous).
    Therefore the equation
    $E[T_{jk}] - E[T_{jk'}] = 0$
    has at most one root in $[0,1]$, and $E[T_{jk}] - E[T_{jk'}]$ is negative below the root and positive above the root.
    As in Theorem~\ref{thm:alg-bounce}, the incentive compatibility region is an intersection of $O(n^2)$ such intervals, corresponding to the set of punishment parameters $b$ where
    $E[T_{jj}] -E[T_{jk}] \geq 0$, which are intervals of the form $[0,b_{jk}]$ or $[b_{jk},1]$.
    As a result, the incentive compatibility region is a single interval $[b^*_{low}, b^*_{high}]$, and we can find $b^*_{low}$ and $b^*_{high}$ explicitly.    
    
    We now need to show, for all $i$ and all $k < k'$, that
    \begin{align}
        \label{eq:ddb-claim}
        \frac{d}{db} E[U_{ik}] \ge \frac{d}{db} E[U_{ik'}].
    \end{align}
    
    We split the argument into three cases:
    $i \le k < k'$, $k < k' < i$, and $k < i \le k'$.
    
    The easiest case to handle is $k < k' < i$,
    where both declared estimates $k$ and $k'$ are underestimates.
    Looking at the mean response time formulas in \cref{lem:measured-trust-formula},
    we see that for any $i > k$,
    $E[U_{ik}]$ is dependent only on $i$, not on $k$.
    In particular, $E[U_{ik}] = E[U_{ik'}]$ in this case.
    As a result, \eqref{eq:ddb-claim} holds as well, because it is an equality.
    
    Next, consider the case where $i \le k < k'$,
    so both declared estimates are accurate or overestimates.
    Looking at \cref{lem:measured-trust-formula},
    we see that the mean response time formulas in this case are relatively simple:
    \begin{align*}
        E[U_{ik}] = \frac{\lambda E[S^2_{\le k}]}{2(1-\rho_{<k})(1-\rho_{\le k})}
        + \frac{z_i}{1-\rho_{< k}}
    \end{align*}
    The formula for $E[U_{ik'}]$ is identical, except that $k$ is replaced by $k'$.
    Differentiating with respect to $b$, we find that
    \begin{align}
        \label{eq:ddb-conservative}
        \frac{d}{db} E[U_{ik}] &=
            \frac{\lambda \frac{d}{db} E[S^2_{\le k}]}{2(1-\rho_{<k})(1-\rho_{\le k})}
            + \frac{\lambda E[S^2_{\le k}] \frac{d}{db} \rho_{<k}}{2(1-\rho_{<k})^2(1-\rho_{\le k})}
            + \frac{\lambda E[S^2_{\le k}] \frac{d}{db} \rho_{\le k}}{2(1-\rho_{<k})(1-\rho_{\le k})^2} \\
            \nonumber
            &+ \frac{z_i \frac{d}{db} \rho_{< k}}{(1-\rho_{< k})^2}
    \end{align}
    Recall that $\rho_{\le k} = \lambda E[S_{\le k}]$.
    
    To prove \eqref{eq:ddb-claim} using \eqref{eq:ddb-conservative},
    we will prove four inequalities about $E[S_{<k}], E[S_{\le k}],$ and $E[S^2_{\le k}]$.
    In each case, the proof is essentially identical, so we focus on $E[S_{\le k}]$.
    The four inequalities for $E[S_{\le k}]$ are:
    \begin{align}
        \label{eq:ddb-reduced}
        E[S_{\le k}] \ge 0, \quad E[S_{\le k}] \le E[S_{\le k'}], \quad \frac{d}{db} E[S_{\le k}] \le 0, \quad
        \frac{d}{db} E[S_{\le k}] \ge \frac{d}{db} E[S_{\le k'}]
    \end{align}
    Plugging these inequalities and the equivalents for $E[S_{< k}]$ and $E[S^2_{\le k}]$ into \eqref{eq:ddb-conservative},
    we find that each term in $\frac{d}{db} E[U_{ik}]$ is negative and closer to zero than
    the equivalent term for $\frac{d}{db} E[U_{ik'}]$. This establishes \eqref{eq:ddb-claim} for this case.
    It therefore suffices to prove \eqref{eq:ddb-reduced}
    and its equivalents.
    $E[S_{\le k}] \ge 0$ simply states that the time a job spends with rank $\le k$ is nonnegative,
    which is immediate. $E[S_{\le k}] \le E[S_{\le k'}]$ states that a job spends less time below smaller ranks than larger ranks, which is also immediate.
    
    For the two claims about $\frac{d}{db}$, we must examine the formula for $S_{\le k}$
    from \cref{sec:measured-trust-formula}, and specifically \eqref{eq:relevant-size}.
    The effect of increasing $b$ on $S_{\le k}$ is to cause jobs to be more likely to be punished sooner,
    moving them from $I_b = 0$ cases to $I_b = 1$ cases.
    This has the effect of decreasing $E[S_{\le k}]$, as desired.
    Moreover, because more jobs spend more time with rank $\le k'$ than with rank $\le k$,
    this effect is more dramatic for $E[S_{\le k'}]$,
    meaning that $\frac{d}{db} E[S_{\le k}] \ge \frac{d}{db} E[S_{\le k'}]$, as desired.
    These statements can be verified through a case analysis on \eqref{eq:relevant-size}.
    This completes the second case.
    
    Finally, we turn to the case where $k < i \le k'$:
    the estimate $k$ is an underestimate, while $k'$ is accurate or an overestimate.%
    \footnote{
        In this case, our argument is specific to MeasuredTrust,
        as opposed to BlindTrust.
        We leave open the question of whether the equivalent result
        for BlindTrust also holds.
    }
    By applying \eqref{eq:ddb-claim} for the previous two cases,
    we can reduce this case to the scenario where $k = i-1$ and $k' = i$:
    
    In particular, to prove that $\frac{d}{db} E[U_{ik}] \ge \frac{d}{db} E[U_{ik'}]$
    for general $k, k'$,
    it suffices to show three steps:
    \begin{align*}
    \frac{d}{db} E[U_{ik}] \ge \frac{d}{db} E[U_{i(i-1)}],\,
    \frac{d}{db} E[U_{i(i-1)}] \ge \frac{d}{db} E[U_{ii}],
    \text{ and }
    \frac{d}{db} E[U_{ii}] \ge \frac{d}{db} E[U_{ik'}].
    \end{align*}
    The first and third are covered by the previous two cases,
    so we only need to prove that $\frac{d}{db} E[U_{i(i-1)}] \ge \frac{d}{db} E[U_{ii}]$
    for all $i$.
    
    To start, let us apply \cref{lem:measured-trust-formula}:
    \begin{align*}
        E[U_{i(i-1)}] &= bE[U_{i(i-1),I_b=1}] + (1-b)E[U_{i(i-1),I_b=0}] \\ 
        [U_{i(i-1),I_b=0}] &= E[U_{ii}] \\
        E[U_{i(i-1)}] - E[U_{ii}] &= bE[U_{i(i-1),I_b=1}] - bE[U_{ii}]
    \end{align*}
    
    To prove \eqref{eq:ddb-claim},
    it suffices to show that
    \begin{align*}
        \frac{d}{db} bE[U_{i(i-1),I_b=1}] \ge \frac{d}{db} bE[U_{ii}].
    \end{align*}
    Let's expand the equations from \cref{lem:measured-trust-formula}:
    \begin{align}
        \label{eq:entire-1}
        bE[U_{i(i-1),I_b=1}] &= \frac{b \lambda E[S^2]}{(1-\rho_{<n+1})(1-\rho)} + \frac{b z_i}{1-\rho}\\
        \label{eq:entire-2}
        bE[U_{ii}] &= \frac{b \lambda E[S_{\le i}^2]}{(1-\rho_{<i})(1-\rho_{\le i})} + \frac{b z_i}{1-\rho_{< i}}
    \end{align}
    First, let us handle the simpler terms on the right (the ``residence time'' terms):
    \begin{align}
        \label{eq:simple-1}
        \frac{d}{db} \frac{b z_i}{1-\rho} &= \frac{z_i}{1-\rho} \\
        \label{eq:simple-2}
        \frac{d}{db} \frac{b z_i}{1-\rho_{< i}} &= \frac{z_i}{1-\rho_{< i}}
        + \frac{b z_i\frac{d}{db}\rho_{< i}}{(1-\rho_{< i})^2}
    \end{align}
    Recall that $\frac{d}{db} \rho_{< i} \le 0$,
    so the second term of \eqref{eq:simple-2} is non-positive.
    Note that $\rho \ge \rho_{< i}$,
    so the desired inequality holds:
    \begin{align*}
        \frac{d}{db} \frac{b z_i}{1-\rho} \ge \frac{d}{db} \frac{b z_i}{1-\rho_{< i}}
    \end{align*}
    
    Now, let us turn to the more complicated terms (the ``queueing time'' terms)
    in \eqref{eq:entire-1} and \eqref{eq:entire-2}.
    First, \eqref{eq:entire-1}:
    \begin{align}
        \label{eq:complicated-1}
        \frac{d}{db} \frac{b \lambda E[S^2]}{(1-\rho_{<n+1})(1-\rho)}
        = \frac{(1-\rho_{<n+1} + b \frac{d}{db} \rho_{<n+1})E[S^2]}{(1-\rho)(1-\rho_{<n+1})^2}
    \end{align}
    Note that $\rho_{<n+1}$ is simply a linear function of $b$,
    as can be seen by examining \eqref{eq:relevant-size},
    from \cref{sec:measured-trust-formula}.
    As a result, there exists some constant $c$ such that
    \begin{align*}
        \rho_{<n+1} = b \frac{d}{db} \rho_{<n+1} + c
    \end{align*}
    Consider the setting where $b=0$:
    no jobs are punished,
    and all jobs run with rank at most $n$.
    Therefore, $\rho_{<n+1}$ is equal to its maximum possible value of $\rho$ when $b=0$.
    Therefore $c = \rho$.
    
    We therefore find that
    \begin{align*}
        1-\rho_{<n+1} + b \frac{d}{db} \rho_{<n+1} = 1- \rho.
    \end{align*}
    
    We use this to simplify \eqref{eq:complicated-1}:
    \begin{align}
        \label{eq:complicated-1a}
        \frac{d}{db} \frac{b \lambda E[S^2]}{(1-\rho_{<n+1})(1-\rho)}
        = \frac{E[S^2]}{(1-\rho_{<n+1})^2}
    \end{align}
    
    Next, let us turn to the first, more complicated term in \eqref{eq:entire-2}.
    We wish to show its derivative is smaller than \eqref{eq:complicated-1a}:
    \begin{align}
        \label{eq:complicated-2}
        \frac{d}{db} \frac{b \lambda E[S_{\le i}^2]}{(1-\rho_{<i})(1-\rho_{\le i})}
        &= \frac{b\frac{d}{db}E[S^2_{\le i}]}{(1-\rho_{<i})(1-\rho_{\le i})} +
        \frac{bE[S^2_{\le i}]\frac{d}{db} \rho_{<i}}{(1-\rho_{<i})^2(1-\rho_{\le i})} +
        \frac{bE[S^2_{\le i}]\frac{d}{db} \rho_{\le i}}{(1-\rho_{<i})(1-\rho_{\le i})^2}\\
        \nonumber
        &+ \frac{E[S^2_{\le i}]}{(1-\rho_{<i})(1-\rho_{\le i})}
    \end{align}
    
    Recall that $\frac{d}{db} E[S^2_{\le i}] \le 0, \frac{d}{db} \rho_{\le i} \le 0,$ and $\frac{d}{db} \rho_{\le i} \le 0$. As a result, the first three terms of \eqref{eq:complicated-2} are negative.
    We therefore have the following bound:
    \begin{align}
        \label{eq:complicated-2a}
        \frac{d}{db} \frac{b \lambda E[S_{\le i}^2]}{(1-\rho_{<i})(1-\rho_{\le i})}
        &\le \frac{E[S^2_{\le i}]}{(1-\rho_{<i})(1-\rho_{\le i})}
    \end{align}
    Compare \eqref{eq:complicated-1a} and \eqref{eq:complicated-2a}.
    Note that $E[S^2] \ge E[S^2_{\le i}]$,
    $\rho_{<n+1} \ge \rho_{<i}$,
    and $\rho \ge \rho_{\le i}$.
    As a result,
    \begin{align*}
        \frac{d}{db} \frac{b \lambda E[S^2]}{(1-\rho_{<n+1})(1-\rho)}
        \ge \frac{d}{db} \frac{b \lambda E[S_{\le i}^2]}{(1-\rho_{<i})(1-\rho_{\le i})}.
    \end{align*}
    
    Putting it all together, we find that $\frac{d}{db} (bE[U_{i(i-1),I_b=1}]) > \frac{d}{db} (bE[U_{ii}])$,
    as desired.
    
    This completes all of the cases, so \eqref{eq:ddb-claim} always holds.
\end{proof}

\subsection{Proof of Theorem 3}
\label{sec:small-noise}

We now prove that for sufficiently accurate estimates, the interval of punishment parameters $b$
for which MeasuredTrust is incentive compatible converges to $[0, 1]$.
We note that our argument here does not extend to BlindTrust (we point out where BlindTrust differs in the proof);
in fact, the statement is false for BlindTrust.\footnote{We provide a simple counterexample for BlindTrust.
Let $S = 1$ with probability $0.99$, otherwise $S=1.1$.
There is no error from estimates, so $M = D_S$.  We take 
$\lambda = 0.8$. For all $b < 0.98$,
BlindTrust is not incentive compatible.
Because BlindTrust does not move a job back in the queue when it is not punished,
there are settings where lying is simply worth the risk, even with fully accurate information.}

\begin{reptheorem}{thm:small-noise}[Small Noise]
    For a given arrival rate $\lambda$ and true size distribution $S$,
    there is some threshold $\epsilon > 0$
    such that for any size-estimate matrix $M$
    compatible with $S$ such that $|M - D_S|_\infty < \epsilon$,
    where $D_S$ is the diagonal, perfectly accurate size-estimate matrix,
    there must exist a punishment probability $b$
    such that MeasuredTrust is incentive compatible.
    Moreover, as $\epsilon \to 0$,
    the interval of $b$ for which MeasuredTrust is incentive compatible
    converges to $[0, 1]$.
\end{reptheorem}
\begin{proof}
    We consider aggressive claims, where the declared estimate $k$ is less than the 
    internal estimate $j$, separately from conservative claims,  where the declared estimate is more than the 
    internal estimate.  
    
    We prove two stronger statements, which together suffice to prove \cref{thm:small-noise}:
    \begin{enumerate}
        \item If $\epsilon$ is sufficiently close to 0,
        then there exists a threshold $b^*_{\epsilon} > 0$ such that for any $b > b^*_{\epsilon}$,
        all aggressive claims are disincentivized.
        Moreover, there exists a sequence of thresholds $b^*_{\epsilon}$ such that $b^*_\epsilon \to 0$ as $\epsilon \to 0$.
        \item If $\epsilon$ is sufficiently close to 0,
        then for any $b$,
        all conservative claims are disincentivized.
    \end{enumerate}

    We start with the first statement.
    We want to show that,
    for sufficiently small $\epsilon$ and $b > b^*_{\epsilon}$,
    for any internal estimate $j$
    and declared estimate $k < j$, $E[T_{jk}] > E[T_{jj}]$
    for all $k < j$.
    
    To prove our result, we will split up the difference
        $E[T_{jk}] - E[T_{jj}]$
    into a negative term, the ``benefit of lying'' and a positive term,
    the ``harm of lying''. We will bound both terms.

    Note that
    \begin{align*}
        E[T_{jk}] = \frac{1}{R_j} \sum_{i} M_{ij} E[U_{ik}].
    \end{align*}
    
    Let us therefore consider $E[U_{ik}]$ for three ranges of values for $i$:
    $i \le k <j$,
    $k < i \le j$,
    and $k < j < i$. 
    
    Using \cref{lem:measured-trust-formula}, the resulting formula is
    \begin{align}
        \nonumber
        E[T_{jk}]- E[T_{jj}] &=
        \textsc{Benefit} + \textsc{Harm} \\
        \textsc{Benefit} &=
        \frac{1}{R_j}\left(\sum_{i \le k < j} M_{ij}(E[U_{ik}] - E[U_{ij}])
        \label{eq:benefit}
        + \sum_{k < i < j} M_{ij} (1-b) (E[U_{ik,I_b=0}] - E[U_{ij}])\right) \\
        \label{eq:harm}
        \textsc{Harm} &= \frac{1}{R_j}\sum_{k < i \le j} M_{ij} b (E[U_{ik,I_b=1}] - E[U_{ij}])
    \end{align}

The \textsc{Benefit} consists only of negative terms, 
while \textsc{Harm} consists only of positive terms, as we explain more fully below.
Note that the interval $i \le k <j$
appears only in \textsc{Benefit},
$k < i \le j$ appears in both \textsc{Benefit} and \textsc{Harm},
and $k < j < i$ appears in neither.

    For $i \le k < j$, 
    there is only a benefit to lying,
    because the smaller declared estimate yields a higher priority service without punishment.

Let us split the case $k < i \le j$
into two scenarios:
$k < i < j$, and $k < i = j$.
For $k < i < j$,
there is both a benefit and a harm to lying:
a benefit if not punished, a harm if punished.
This is reflected in \eqref{eq:benefit} and \eqref{eq:harm}.
    
    Note however that if $k < i = j$,
    there is no benefit to lying,
    even if the job is not punished,
    because the job will finish by running at rank $i = j$
    in either case.
    Formally, $E[U_{ik,I_b=0}] = E[U_{ii}]$,
    as can be seen by comparing \eqref{eq:measured-trust-unbounced} and \eqref{eq:measured-trust-safe}
    in \cref{lem:measured-trust-formula}.%
\footnote{This is the step of the argument that holds for MeasuredTrust
but not for BlindTrust. If $k < i = j$,
there is a benefit to lying under BlindTrust,
because if the job is not punished,
the lying mean response time $E[U^{BT}_{ik,I_b=0}]$
is smaller than the honest mean response time $E[U^{BT}_{ii}]$.
See \cref{lem:blind-trust-formula}.}
As a result, $k < i = j$ appears only in \eqref{eq:harm}.

    Finally, for $k < j < i$,
    lying has no effect on mean response time, as can be seen by looking at \cref{lem:measured-trust-formula}.
    This is one of the key results of the SOAP approach, described in \cref{sec:formulas}.
    Intuitively, this is because the job is punished with the same probability 
    for both the declared estimate and internal estimate, and it doesn't matter when it is punished.  

    Now, we need to bound the benefit of lying close to zero,
    and bound the harm of lying away from zero.
    Specifically, we first show that
    there exists a constant $c_1$
    dependent on $\lambda$ and $S$ but not on $b$ and $\epsilon$
    such that $|\textsc{Benefit}| \le c_1 \epsilon.$
    
    In the other direction,
    we show that
    there exists a constant $c_2 > 0$ dependent on $\lambda$ and $S$ but not on $b$ and $\epsilon$
    such that,
    for sufficiently small $\epsilon$, $\textsc{Harm} \ge c_2 b.$
    
    From these bounds on \textsc{Benefit} and \textsc{Harm},
    the desired result holds with $b^*_\epsilon = \frac{c_1 \epsilon}{c_2}$.
  
    To prove both bounds,
    we make use of our assumption that $|M - D_S|_\infty < \epsilon$.
    Specifically, this means that $M_{ij} < \epsilon$
    for all $i \neq j$,
    and $M_{ii} \ge P(S = z_i) - \epsilon$ for all $i$.
    
    Note that in \eqref{eq:benefit},
    every term includes an $M_{ij}$
    for some $i \neq j$.
    As a result, we can bound the benefit:
    \begin{align*}
        |\textsc{Benefit}| \le \frac{\epsilon}{R_j} \left( \sum_{i \le k,j > k} (E[U_{ik}] - E[U_{ij}]) + \sum_{j > k, i \in (k, j)} (1-b) (E[U_{ik,I_b=0}] - E[U_{ij}]) \right)
    \end{align*}
    Now, we merely need to bound $E[U_{ik}]$ by a value dependent only on $S$ and $\lambda$.
    Looking at the expressions for mean response time given in \cref{sec:measured-trust-formula},
    we can define a simple bound, $u^*$:
    \begin{align*}
        u^* &:= \frac{\lambda E[S^2]}{2(1-\rho)^2} + \frac{z_n}{1-\rho} \\
        \forall i, k, \quad E[U_{ik}] &\le u^*, \quad E[U_{ik, I_b = 0}] \le u^*.
    \end{align*}
    Note that $E[S^2] \ge E[S^2_{\le i}]$ for any $i$,
    and likewise $\rho \ge \rho_{\le i}$ for any $i$.
    
    As a result, we can give a simple bound on the benefit of lying only dependent on $S, \lambda,$ and $\epsilon$.
    \begin{align*}
        |\textsc{Benefit}| \le \epsilon \frac{n u^*}{R_j}.
    \end{align*}
    
    Next, we need to lower bound the harm of lying.
    In the expression for the harm of lying given in \eqref{eq:harm},
    let us focus on giving a lower bound on the term where $i = j$.
    \begin{align*}
        \textsc{Harm} &\ge \frac{M_{jj}}{R_j} b (E[U_{jk, I_b=1}] - E[U_{jj}]) \\
        & \ge \frac{(P(S = z_j) - \epsilon)}{R_j} b (E[U_{jk, I_b=1}] - E[U_{jj}])
    \end{align*}
    
    For $\epsilon < P(S = z_j)$,
    to prove that $\textsc{Harm} \ge c_2 b$,
    it suffices to show that $E[U_{jk, I_b=1}] - E[U_{jj}] > 0$.
    
    From \cref{lem:measured-trust-formula},
    let us examine the expressions for $E[U_{jk, I_b=1}]$ and $E[U_{jj}]$.
    \begin{align*}
        E[U_{jj}] &=
        \frac{\lambda E[S^2_{\le j}]}{2(1-\rho_{<j})(1-\rho_{\le j})}
        + \frac{z_j}{1-\rho_{<j}} \\
        E[U_{jk,I_b=1}] &=
        \frac{\lambda E[S^2]}{2(1-\rho_{<n+1})(1-\rho)}
        + \frac{z_j}{1-\rho}  
    \end{align*}
    Note that $\rho_{\le j} \le \rho_{<n+1}$, and that $E[S^2_{\le j}] \le E[S^2]$.
    Therefore, to prove a separation between $E[U_{jk, I_b=1}]$ and $E[U_{jj}]$,
    it suffices to prove a separation between $\rho_{<j}$ and $\rho$,
    or equivalently between $E[S_{<j}]$ and $E[S]$.
    
    Let us focus on the contribution to $E[S]$ and $E[S_{<j}]$ of jobs of class $j$.
    First, from \cref{sec:measured-trust-formula},
    let us write the formula for $E[S_{<j}]$:
    \begin{align*}
        E[S_{<j}] = \sum_{i' \in [n], j' \in [n]} M_{i'j'}E[S_{i'j', \le j-1}]
    \end{align*}
    For $E[S]$,
    the term $S_{i'j', \le j-1}$
    is replaced by $z_{i'}$, the entire size of the job.
    By definition, $E[S_{i'j', \le j-1}]$,
    the amount of service the job receives while having rank $\le j-1$,
    is at most $z_{i'}$.
    
    Therefore, we only need to prove strict inequality in some case which occurs with positive probability.
    Let us focus on the case $i' = j$, which occurs with positive probability $P(S = z_j)$.
    Using \eqref{eq:relevant-size}, the values of $S_{j j', \le j-1}$ simplify to:
    
    \begin{align*}
        S_{j j', \le j-1} = \begin{cases}
            j' > j-1 & 0 \\
            j' \le j-1 \land I_b = 1 & z_{j'} \\
            j' \le j-1 \land I_b = 0 & z_{j-1}
        \end{cases}
    \end{align*}
    In all three cases, $S_{j j', \le j-1} \le z_{j-1}$.
    This is strictly smaller than $z_j$,
    demonstrating the desired separation.
    
    We can therefore conclude that $E[S] > E[S_{<j}]$,
    and hence that $E[U_{jk, I_b=1}] > E[U_{jj}]$,
    and hence that $\textsc{Harm} \ge c_2 b$
    for some $c_2 > 0$, as desired.
    
    This completes the proof for aggressive claims, where $k < j$.
    
    We now turn to conservative claims, where $k > j$.
    Now, we wish to show that for sufficiently small $\epsilon$,
    $E[T_{jk}] > E[T_{jj}]$.
    Again, we split the difference $E[T_{jk}] - E[T_{jj}]$ into terms 
    according to whether lying is beneficial or harmful.
    
    We again consider three ranges of values for $i$:
    $i \le j < k$,
    $j < i \le k$,
    $j < k < i$.
    Using \cref{lem:measured-trust-formula},
    the resulting formula is
    \begin{align*}
        E[T_{jk}] - E[T_{jj}] &= \textsc{Benefit} + \textsc{Harm} \\
        \textsc{Benefit} &= \frac{1}{R_j} \sum_{j < i \le k} M_{ij} b (E[U_{ik}] - E[U_{ij,I_b=1}]) \\
        \textsc{Harm} &= \frac{1}{R_j} \left(\sum_{i \le j < k} M_{ij} (E[U_{ik}] - E[U_{ij}]) +
             \sum_{j < i < k} M_{ij} (1-b) (E[U_{ik}] - E[U_{ij,I_b=0}]) \right)
    \end{align*}
    This formula is identical to \eqref{eq:benefit} and \eqref{eq:harm},
    swapping $j$ and $k$, as well as \textsc{Benefit} and \textsc{Harm}.
    
    As before, every term in \textsc{Benefit}
    includes some $M_{ij}$ where $i \neq j$.
    As a result, $|\textsc{Benefit}| \le \epsilon \frac{n u^*}{R_j}$,
    as a conservative bound.
    
    As for harm, let us again focus on the term where $i = j$:
    \begin{align*}
        \textsc{Harm} \ge \frac{M_{jj}}{R_j} (E[U_{jk}] - E[U_{jj}]).
    \end{align*}
    It therefore suffices to prove that $E[U_{jk}] > E[U_{jj}]$.
    Looking at \cref{lem:measured-trust-formula},
    we merely need to prove that $E[S_{\le k}] > E[S_{\le j}]$,
    which again follows from a simple case analysis of $S_{ij,\le \ell}$
    given in \eqref{eq:relevant-size}.
    
    Therefore there exists an $\epsilon > 0$
    such that for any $b$,
    $E[T_{jk}] > E[T_{jj}]$ for all $k > j$, as desired.
\end{proof}
\section{Empirical Validation}
\label{sec:empirical}

\begin{figure}
    \centering
    \includegraphics[width=0.7\textwidth]{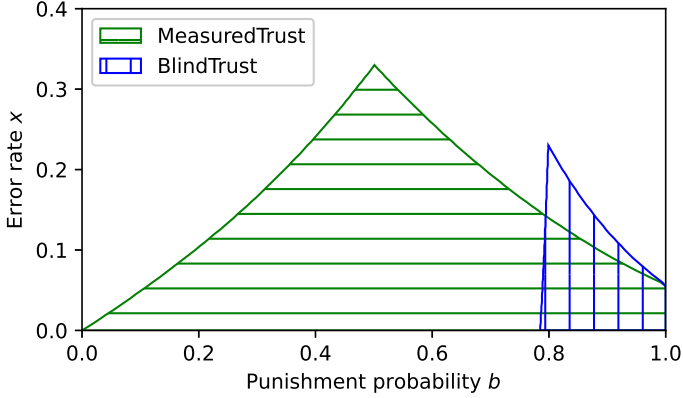}
    \caption{Incentive compatible region under MeasuredTrust and BlindTrust policies.
    Under each policy, there is a single interval of incentive compatible punishment parameters $b$ for any error rate $x$, or no such $b$.
    That interval is largest for $x=0$, no error, and shrinks monotonically for larger $x$.
    Job size distribution is 0.4 w.p. $\frac{1}{2},$ 0.8 w.p. $\frac{1}{4}$, 1.6 w.p. $\frac{1}{8}$, 3.2 w.p. $\frac{1}{8}$. $E[S]= 1$, $\lambda = \rho = 0.8$.
    Errors are uniform: correct estimate with probability $1-x$,
    each other possible value as the estimate with probability $x/3$.
    This chart was generated using a step size of $0.005$ for $x$ and $0.001$ for $b$.
}
    \label{fig:incentive-compat}
\end{figure}

We have proven strong theoretical results about the MeasuredTrust policy.
In \cref{thm:single-interval}, we showed that for any $M$ and $\lambda$,
the set of punishment parameters $b$ for which MeasuredTrust is incentive compatible
forms a single connected interval.
In \cref{thm:small-noise}, we showed that this interval expands to $[0, 1]$
in the limit as the noise vanishes.

To illustrate and empirically validate our theoretical results,
we make use of the SOAP formulas from \cref{sec:formulas}
to plot the regions for which MeasuredTrust and BlindTrust are incentive compatible.
We also empirically validate the social benefit of MeasuredTrust and BlindTrust.

In \cref{fig:incentive-compat},
we consider a setting with estimate noise parameterized by an error rate $x$,
where the estimate is incorrect with probability $x$, independent of the true size.  When an estimate is incorrect, it takes on one of the other three values uniformly at random.  
We plot the region of incentive compatibility with respect to the error rate $x$
and the punishment parameter $b$ for both MeasuredTrust and BlindTrust.

Notice that for all error rates $x$,
the set of punishment parameters $b$ for which MeasuredTrust is incentive compatible
forms a single interval, if any such $b$ exists.
This matches our result in \cref{thm:single-interval}.
In this example,
BlindTrust also has the single-interval property. We leave proving this generally or finding a counterexample to future work.

Notice also that for error rates $x$ near $0$,
the set of incentive-compatible $b$ under MeasuredTrust converges to the entire interval
$[0, 1]$, matching our result in \cref{thm:small-noise}.
The same does not hold for BlindTrust, showing that this result is specific to MeasuredTrust.

Another important behavior illustrated by \cref{fig:incentive-compat}
is the maximum error rate $x$ for which there exists an incentive compatible punishment parameter $b$.
For MeasuredTrust, this maximum error rate is $x=0.33$, $b=0.501$,
while for BlindTrust, this maximum error rate is $x=0.23$, $b = 0.799$.
As a result, MeasuredTrust is robust to more error than BlindTrust in this setting,
if the punishment parameter is tuned appropriately.

Now, let us turn to the overall mean response times under these policies.
In particular, let us examine the optimal mean response time
achievable while maintaining incentive compatibility in this setting.

\begin{figure}
    \centering
    \includegraphics[width=0.7\textwidth]{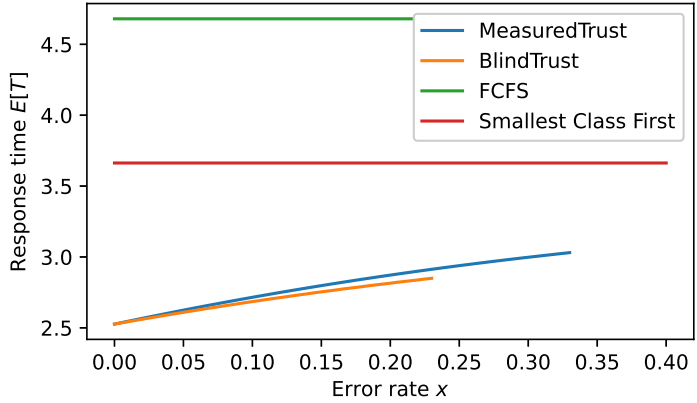}
    \caption{Mean response time under MeasuredTrust and BlindTrust policies,
    as compared to two blind policies: First-Come First-Served; and Smallest Class First. Punishment parameter $b$ selected to minimize mean response time, while maintaining incentive compatibility.
    Both estimate-aware policies are socially beneficial compared to the blind policies at all $x$ for which incentive compatibility is possible.
    Same setting as \cref{fig:incentive-compat}.}
    \label{fig:response}
\end{figure}

In \cref{fig:response}, we plot such optimal mean response times
for MeasuredTrust and BlindTrust, in the same setting as \cref{fig:incentive-compat}.
The two policies have very similar mean response times,
increasing mildly as the error rate $x$ increases.

We also plot the mean response time of two blind policies,
which do not consider estimate information:
First-Come First-Served (FCFS), and Smallest Class First.
FCFS simply serves jobs in the order they arrive.
Smallest Class First (SCF) attempts to serve the job of smallest class,
given its limited information.
To do so, it checks whether any job might have true size $z_1$.
If so, such a job is served, with FCFS tiebreaking.
Otherwise, SCF checks whether any job might have true size $z_2$,
and so on.

From \cref{fig:response},
we see that for any error rate $x$ for which incentive compatibility
can be achieved, both estimate-aware policies achieve significantly better
mean response times than the two blind policies,
achieving significant social benefit by making use of estimates.  Of course, this is the goal when using estimates -- to achieve better scheduling performance for the whole of the system. 

\section{Conclusion}

We have provided a theoretical framework for studying incentive-compatible scheduling algorithms for single queues, and examined some natural algorithms in this setting.  We view this as an opening to a wide array of questions.  For example, extending the framework to more complex systems, such as multi-server systems, would be a natural step.  As an initial question, one could try to extend the results to $M/G/k$-type queues, where there are $k$ servers and each job runs on only one server at a time.  More challenging settings would consider malleable jobs that can run in parallel on multiple servers.

Another direction for future work would be to characterize
the amount of correlation between size and estimate
necessary for incentive compatibility and social benefit to be achievable.
Intuitively, if the estimates are noisy enough,
a blind policy might be better,
while benefit should be achievable for less noisy estimates.

Finally, we have chosen to work with a simplified system where job times and estimates come from a finite set.  There appear to be various technical challenges in generalizing to continuous distributions;  we leave this as a problem for future work.  

\begin{acks}
Isaac Grosof was supported in part by NSF grants CMMI-1938909 and CSR-1763701.
Michael Mitzenmacher was supported in part by NSF grants CCF-2101140, CNS-2107078, and DMS-2023528, and by a gift to the Center for Research on Computation and Society at Harvard University.
\end{acks}

\bibliographystyle{plain}
\bibliography{bibfile}

\eat{
\section{Notes}
Q: If perfectly accurate estimates, can we construct a counter-example to social benefit?
A: Yes: Sizes 50\% 1, 50\% 1.01, $\lambda = 0.5$.
Not socially beneficial relative to FCFS.
If wanted to always get social benefit relative to all policies,
you'd have to run some flavor of ``estimate SRPT'',
and to achieve with $\epsilon$ error, want something robust like
``estimate SRPT-B'' or something similar. Future work.

Thoughts to do:  compare with monotonic rank policies, such as monotonic SRPT or Gittins.  

Next things to work on:

Isaac: Equations (SOAP) section

Michael: Theorem 1 \& 2

Isaac: Theorem 3

Empirical.
}

\end{document}